\documentclass[conference]{IEEEtran}
\IEEEoverridecommandlockouts
\usepackage{cite}
\usepackage[colorlinks,urlcolor=blue,linkcolor=red,anchorcolor=red,citecolor=green]{hyperref}
\usepackage{amsmath,amssymb,amsfonts}
\usepackage{algorithmic}
\usepackage{graphicx}
\usepackage{textcomp}
\usepackage{xcolor}
\usepackage{marvosym}
\usepackage{multirow}
\usepackage{makecell}
\def\BibTeX{{\rm B\kern-.05em{\sc i\kern-.025em b}\kern-.08em
    T\kern-.1667em\lower.7ex\hbox{E}\kern-.125emX}}

\begin{document}

\title{A Semi-Supervised Approach with Error Reflection for Echocardiography Segmentation}

\DeclareRobustCommand*{\IEEEauthorrefmark}[1]{%
  \raisebox{0pt}[0pt][0pt]{\textsuperscript{\footnotesize #1}}%
}

\author{
    \IEEEauthorblockN{
        Xiaoxiang Han\IEEEauthorrefmark{1,2,$\dagger$}, 
        Yiman Liu\IEEEauthorrefmark{3,4,$\dagger$}
        \thanks{$\dagger$ Xiaoxiang Han and Yiman Liu contribute equally to this work, and are co-first authors.
},
        Jiang Shang\IEEEauthorrefmark{1,2},
        Qingli Li\IEEEauthorrefmark{5},
        Jiangang Chen\IEEEauthorrefmark{5},
        Menghan Hu\IEEEauthorrefmark{5},\\
        Qi Zhang\IEEEauthorrefmark{1,2,\Letter}
        \thanks{\Letter{} Corresponding authors: Qi Zhang, Yuqi Zhang and Yan Wang.},
        Yuqi Zhang\IEEEauthorrefmark{3,4,\Letter},
        Yan Wang\IEEEauthorrefmark{5,\Letter}
        \thanks{Email: hanxx@shu.edu.cn, liuyiman@scmc.com.cn, zhangq@t.shu.edu.cn, zhangyuqi@scmc.com.cn, ywang@cee.ecnu.edu.cn}
    }
    \IEEEauthorblockA{
        \IEEEauthorrefmark{1} \textit{School of Communication and Information Engineering, Shanghai University, Shanghai, P.R.China}\\
        \IEEEauthorrefmark{2} \textit{The SMART (Smart Medicine and AI-based Radiology Technology) Lab,}\\
        \textit{Shanghai Institute for Advanced Communication and Data Science, Shanghai University, Shanghai, P.R.China}\\
        \IEEEauthorrefmark{3} \textit{Department of Pediatric Cardiology, Shanghai Children’s Medical Center,}\\ \textit{School of Medicine, Shanghai Jiao Tong University, Shanghai, P.R.China}\\
        \IEEEauthorrefmark{4} \textit{Shanghai Engineering Research Center of Intelligence Pediatrics (SERCIP), Shanghai, P.R.China}\\
        \IEEEauthorrefmark{5} \textit{Shanghai Key Laboratory of Multidimensional Information Processing,}\\
        \textit{East China Normal University, Shanghai, P.R.China}
    }
}

\maketitle

\begin{abstract}
Segmenting internal structure from echocardiography is essential for the diagnosis and treatment of various heart diseases. Semi-supervised learning shows its ability in alleviating annotations scarcity. While existing semi-supervised methods have been successful in image segmentation across various medical imaging modalities, few have attempted to design methods specifically addressing the challenges posed by the poor contrast, blurred edge details and noise of echocardiography. These characteristics pose challenges to the generation of high-quality pseudo-labels in semi-supervised segmentation based on Mean Teacher. Inspired by human reflection on erroneous practices, we devise an error reflection strategy for echocardiography semi-supervised segmentation architecture. The process triggers the model to reflect on inaccuracies in unlabeled image segmentation, thereby enhancing the robustness of pseudo-label generation. Specifically, the strategy is divided into two steps. The first step is called reconstruction reflection. The network is tasked with reconstructing authentic proxy images from the semantic masks of unlabeled images and their auxiliary sketches, while maximizing the structural similarity between the original inputs and the proxies. The second step is called guidance correction. Reconstruction error maps decouple unreliable segmentation regions.  Then, reliable data that are more likely to occur near high-density areas are leveraged to guide the optimization of unreliable data potentially located around decision boundaries. Additionally, we introduce an effective data augmentation strategy, termed as multi-scale mixing up strategy, to minimize the empirical distribution gap between labeled and unlabeled images and perceive diverse scales of cardiac anatomical structures. Extensive experiments on a public echocardiography dataset CAMUS, and a private clinical echocardiography dataset demonstrate the competitiveness of the proposed method.
\end{abstract}

\begin{IEEEkeywords}
Semi-supervised learning, Echocardiography, Image segmentation
\end{IEEEkeywords}

\section{Introduction}
Transthoracic echocardiography (TTE) possesses characteristics of non-invasiveness, absence of radiation, and cost-effectiveness, rendering it widely applicable in clinical settings. TTE aids the medical team in discerning the nature and extent of lesions in complex congenital heart diseases (CHD)~\cite{hillis2005basic}. Segmenting the internal structures from echocardiography can aid clinicians in identifying various cardiac conditions and assessing cardiac functionality. With the advancement of deep learning~\cite{he2016deep,liu2023edmae}, significant advancements have been made in medical image segmentation~\cite{ronneberger2015u,chen2021transunet}. Since then, this technology has been introduced into echocardiography segmentation~\cite{leclerc2019deep,painchaud2022echocardiography} and achieves promising results. However, training an accurate echocardiography segmentation model usually requires a large amount of labeled data. Data annotation is a time-consuming and labor-intensive task, especially when it comes to annotating medical images, which requires domain-specific knowledge. Semi-supervised learning (SSL)~\cite{bai2017semi} has shown great potential to alleviate annotations scarcity, which typically obtains high-quality segmentation results by learning from a limited amount of labeled data and a large set of unlabeled data directly. 

\begin{figure}[t]
    \includegraphics[width=0.5\textwidth]{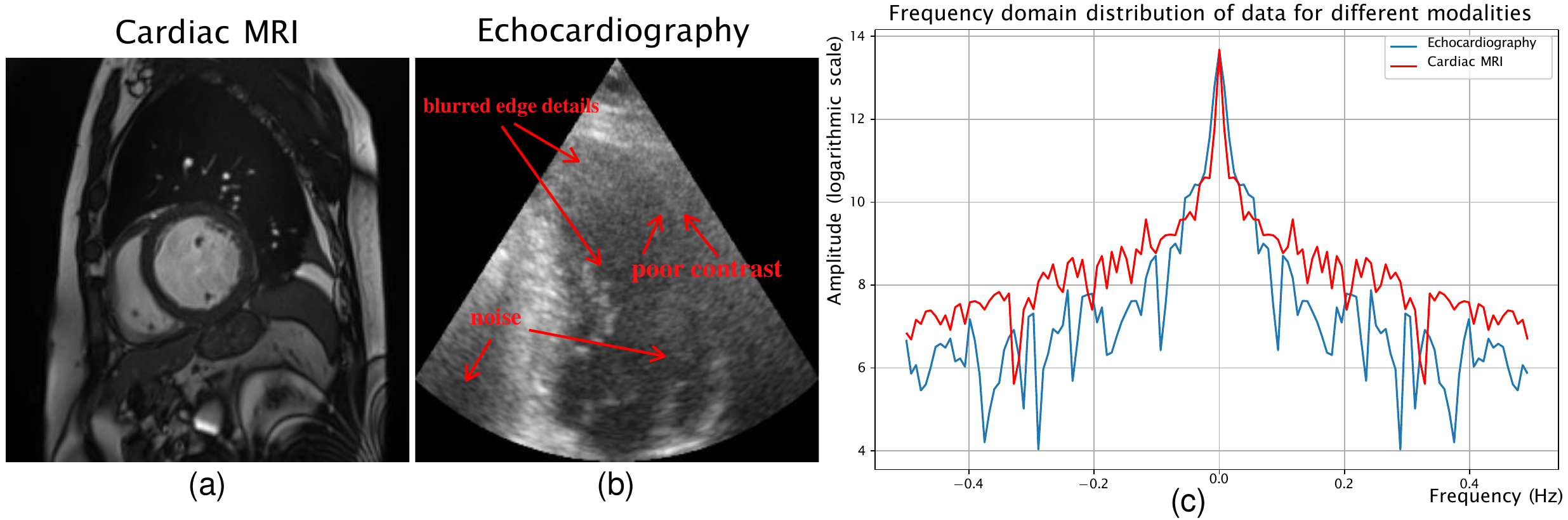}
    \caption{Comparison between echocardiography and cardiac MRI. (a) and (b) present a cardiac MR image and an echocardiographic image, respectively. Visually, echocardiography exhibits poor contrast, blurred edge details and more noise compared to cardiac MRI. (c) depicts their spectra, revealing fewer high-frequency components in echocardiography.} \label{frequency}
\end{figure}

Many recently successful SSL methods typically leverage unlabeled data by performing unsupervised consistency regularization. These approaches base on the smoothness assumption, enforcing consistent predictions through perturbations at the image level~\cite{yu2019uncertainty,bai2023bidirectional}, feature level~\cite{ouali2020semi}, or network level~\cite{wu2021semi}. In addition, DTC~\cite{luo2021semi} established task-level regularization, and SASSnet~\cite{li2020shape} introduced adversarial loss to learn consistent shape representations from the dataset. Image-level data perturbation or augmentation is common yet potent, because pixels in the same map share semantics and are closer, such as CutMix~\cite{french2019semi}, ClassMix~\cite{olsson2021classmix}, and ComplexMix~\cite{chen2021complexmix}. BCP~\cite{bai2023bidirectional} focused on designing consistent learning strategies for both labeled and unlabeled data. Some studies attempted to introduce semi-supervised methods into echocardiography segmentation~\cite{madani2018deep,guo2023improved}. Zhao et al.~\cite{zhao2024boundary} employed boundary attention and feature consistency constraints to improve the performance of semi-supervised echocardiography segmentation. Wu et al.~\cite{wu2022semi} proposed an adaptive spatiotemporal semantic alignment method for semi-supervised echocardiography video segmentation.

Although existing semi-supervised methods have made significant advancements in image segmentation across various medical imaging modalities, they fail to fully tailor the design to the characteristics of echocardiography. As depicted in Fig.\ref{frequency}, spectral analysis comparing echocardiographic and cardiac MR images reveals a reduced presence of high-frequency components in echocardiography. The high-frequency components in an image typically convey information such as edges, fine details, and textures. Therefore, whether from visual perception or spectral analysis, echocardiography typically exhibits poor contrast, blurred edge details and more noise compared to other medical imaging modalities. These characteristics pose challenges to the generation of high-quality fine pseudo-labels from unlabeled images by the teacher network. In the Mean Teacher~\cite{tarvainen2017mean} architecture, the teacher network generates pseudo-labels for the unlabeled images to guide the student network, thereby emphasizing the importance of producing high-quality pseudo-labels.

Therefore, we devise an \textbf{error reflection strategy} inspired by human intuition to reflect upon and rectify erroneous practices. When humans encounter erroneously segmented outputs, they may reflect upon the original image and attempt to correct segmentation. Specifically, the strategy is divided into two steps. The first step is called \textbf{reconstruction reflection}. The network is tasked to reconstruct authentic proxy images from the semantic masks of unlabeled images and their auxiliary sketches. If the segmentation is accurate, the reconstructed proxy shall exhibit a high degree of similarity to the original image. Otherwise, the proxy shall manifest deficiencies. The second step is called \textbf{guidance correction}. The error map, generated by discrepancies between original and proxy images, decouples unlabeled image segmentation into reliable and unreliable areas, guiding optimization of potentially unreliable data near decision boundaries with reliable data from high-density regions. The strategy encourages the model to reflect on itself to identify deficiencies in pseudo-labels, thereby enhancing the robustness of pseudo-label generation. In addition, data augmentation in semi-supervised learning is important. We introduce an effective data augmentation strategy, termed as the \textbf{multi-scale mixing up strategy}. Specifically, we improve the strategy of partitioning labeled and unlabeled images into small, variable-sized patches akin to a puzzle, by randomly mixing them while preserving relative positions and varying patch sizes. This facilitates the minimization of the empirical distribution gap between labeled and unlabeled images while enabling the perception of various scales of cardiac anatomical structures. We verify the proposed method on the popular public echocardiography dataset CAMUS~\cite{leclerc2019deep} and a private clinical echocardiography dataset. Extensive experiments demonstrate the advantages and competitiveness of our method. Our approach notably does not modify the U-Net architecture.

\begin{figure*}[t]
    \centering
    \includegraphics[width=0.8\textwidth]{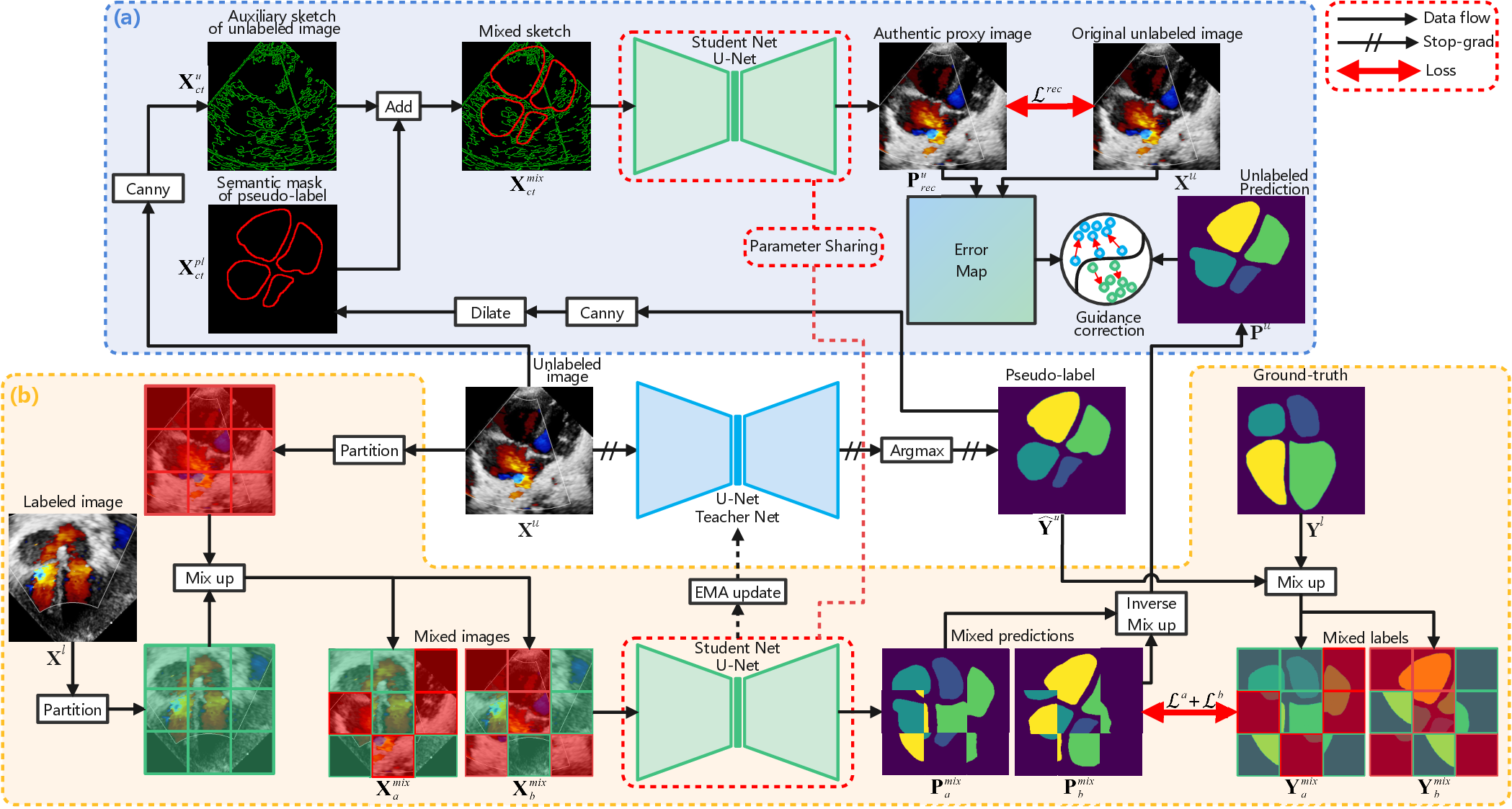}
    \caption{The pipeline of the proposed approach. (a) Error reflection strategy, including the reconstruction reflection step and the guidance correction step. The lines in $\mathbf{X}^u_{ct}$, $\mathbf{X}^{pl}_{ct}$, and $\mathbf{X}^{mix}_{ct}$ in the figure are pseudo-colored, with red lines derived from the semantic masks of pseudo-labels and green lines from the auxiliary sketches of unlabeled images. (b) Multi-scale mixing up (data augmentation) strategy. The patches marked red in the figure are from unlabeled images, while those marked green are from labeled images. \textbf{Note:} (a) All student networks (green ones) are the same network with shared parameters. (b) The figure shows \textbf{only the case where N=3.}} \label{refection}
\end{figure*}

\section{Method}

We define the RGB image of an echocardiography scan as $\mathbf{X}\in{\mathbb{R}^{W\times{H\times{C}}}}$, where $W$, $H$, and $C$ respectively represent the width, height, and channels of the image. The goal of semi-supervised segmentation is to predict the semantic label of each pixel $k\in{\mathbf{X}}$, which constitutes into a prediction label map $\widehat{\mathbf{Y}}\in{\left \{ 0,1,...,K \right \}^{W\times{H}}}$, where $K=0$ represents back-ground, and $K>0$ indicates the cardiac anatomy class. The training set $\mathcal{D}$ comprises two subsets of different sizes: $\mathcal{D}=\mathcal{D}^l\cup{\mathcal{D}^u}$, where $\mathcal{D}^l=\left \{ (\mathbf{X}^l_i,\mathbf{Y}^l_i) \right \}^N_{i=1}$ and $\mathcal{D}^u=\left \{ \mathbf{X}^u_i \right \}^{M+N}_{i=N+1}$ ($N\ll{M}$). The pipeline of the proposed approach is shown in Fig.\ref{refection}, based on Mean Teacher architecture~\cite{tarvainen2017mean}. The proposed framework consists of the error reflection strategy and the multi-scale mixing up strategy. When training the framework, a mini-batch $\mathcal{B}$ consists of $n$ images $\mathbf{X}^\mathcal{B}\in{\mathbb{R}^{n\times{W\times{H\times{C}}}}}$. Assuming $n=2$, each $\mathcal{B}$ comprises $1$ labeled image $\mathbf{X}^l\in{\mathbb{R}^{W\times{H\times{C}}}}$ and $1$ unlabeled image $\mathbf{X}^u\in{\mathbb{R}^{W\times{H\times{C}}}}$, sampled from sets $\mathcal{D}^l$ and $\mathcal{D}^u$ respectively.

\subsection{Error reflection strategy}
\subsubsection{Reconstruction reflection step}
The workflow of the reconstruction reflection step is as follows: Unlabeled image $\mathbf{X}^u$ is fed into the teacher network $\mathcal{F}_t(\cdot;\mathbf{\Theta}_t)$ to obtain prediction map $\mathbf{P}^u_t\in{\mathbb{R}^{K\times{W\times{H}}}}$, where $K$ represents the number of classes. $\mathbf{P}^u_t$ is processed by argmax function $f_{argmax}(\cdot)$ to obtain pseudo-label mask $\mathbf{\widehat{Y}}^u\in{\left \{ 0,1,...,K \right \}^{W\times{H}}}$. The sketches $\mathbf{X}^u_{ct},\mathbf{X}^{pl}_{ct}\in{\left \{ 0,1 \right \}^{W\times{H}}}$ are extracted by the Canny operator $f_{canny}(\cdot)$ from the unlabeled image $\mathbf{X}^u$ and its pseudo-label mask $\mathbf{\widehat{Y}}^u$. Then, $\mathbf{X}^{pl}_{ct}$ is bolded by the dilation morphological operation $f_{dilate}(\cdot)$. Afterwards, $\mathbf{X}^u_{ct}$ and $\mathbf{X}^{pl}_{ct}$ are merged into a new sketch $\mathbf{X}^{mix}_{ct}\in{\left \{ 0,1 \right \}^{W\times{H}}}$ by the addition operation $f_{add}(\cdot)$. The mixed sketch $\mathbf{X}^{mix}_{ct}$ is sequentially processed by the student network $\mathcal{F}_s(\cdot;\mathbf{\Theta}_s)$ to reconstruct the authentic proxy image $\mathbf{P}^u_{rec}\in{\mathbb{R}^{W\times{H\times{C}}}}$. The whole process can be expressed succinctly as:
\begin{equation}
\mathbf{\widehat{Y}}^u = f_{argmax}(\mathcal{F}_t(\mathbf{X}^u;\mathbf{\Theta}_t))),
\end{equation}
\begin{equation}
\mathbf{X}^u_{ct} = f_{canny}(\mathbf{X}^u), \mathbf{X}^{pl}_{ct} = f_{dilate}(f_{canny}(\mathbf{\widehat{Y}}^u)),
\end{equation}
\begin{equation}
\mathbf{P}^u_{rec} = \mathcal{F}_s(f_{add}(\mathbf{X}^u_{ct}, \mathbf{X}^{pl}_{ct});\mathbf{\Theta}_s).
\end{equation}

The calculation of structural similarity (SSIM) loss for the reconstruction task can be denoted as: $\mathcal{L}^{rec} = \ell_{ssim}(\mathbf{P}^u_{rec},\mathbf{X}^u)$, where the unlabeled image $\mathbf{X}^u$ serves as the ground truth for $\mathbf{P}^u_{rec}$. $\ell_{ssim}(\cdot)$ denotes the SSIM loss function.

\subsubsection{Guidance correction step}
First, the pixel-wise error map $\mathbf{M}^{err}\in{\mathbb{R}^{W\times{H}}}$ between the reconstructed proxy image $\mathbf{P}^u_{rec}$ and the unlabeled original image $\mathbf{X}^u$ is computed. As the error decreases continuously during the training process, a fixed error threshold is inappropriate. We devise a simple dynamic error threshold based on $\mathbf{M}^{err}$ to obtain the unreliable segmentation region map $\mathcal{M}^{ur}\in{\left \{ 0,1 \right \}^{W\times{H}}}$. The process can be expressed succinctly as:
\begin{equation}
\mathbf{M}^{err} = f_{abs}(f_{norm}(\mathbf{P}^u_{rec}) - f_{norm}(\mathbf{X}^u)),
\end{equation}
\begin{equation}
\mathcal{M}^{ur} = \mathbf{M}^{err} > f_{max}(\mathbf{M}^{err}) / 2,
\end{equation}
where $f_{norm}(\cdot)$ denotes normalization, $f_{abs}(\cdot)$ represents taking the absolute value, and $f_{max}(\cdot)$ indicates obtaining the maximum value.

The unlabeled image prediction $\mathbf{P}_s^u\in{\mathbb{R}^{K\times{W\times{H}}}}$ is obtained by mixed predictions $\mathbf{P}^{mix}_a$ and $\mathbf{P}^{mix}_b$ (their generation process is shown in next subsection) through the inverse mixing up operation $\mathcal{O}_{mix}^{inverse}(\cdot)$. Then, $\mathbf{P}_s^u$ and $\mathbf{P}_{t}^u$ are respectively multiplied by $\mathcal{M}^{ur}$ point-by-point to obtain their unreliable prediction regions $\mathbf{P}_{s,ur}^u,\mathbf{P}_{t,ur}^u\in{\mathbb{R}^{K\times{W\times{H}}}}$. Subsequently, the pixel-wise values of $\mathbf{P}_{t,ur}^u$ and $\mathbf{P}_{s,ur}^u$ are compared, and a new mask $\mathcal{M}^{g}\in{\left \{ 0,1 \right \}^{W\times{H}}}$ is generated when the predicted probability of $ \mathbf{P}_{t,ur}^u$ is greater than that of $\mathbf{P}_{s,ur}^u$. The process is denoted as:
\begin{equation}
\mathbf{P}_s^u = \mathcal{O}_{mix}^{inverse}(\mathbf{P}^{mix}_a,\mathbf{P}^{mix}_b),
\end{equation}
\begin{equation}
\mathbf{P}_{s,ur}^u = \mathbf{P}_s^u \odot \mathcal{M}^{ur}, \mathbf{P}_{t,ur}^u = \mathbf{P}_t^u \odot \mathcal{M}^{ur},
\end{equation}
\begin{equation}
\mathcal{M}^{g} = \mathbf{P}_{t,ur}^u > \mathbf{P}_{s,ur}^u,
\end{equation}
where $\odot$ denotes point-by-point multiplication.

For the unreliable segmented regions $\mathbf{P}_{t,ur}^u$ and $\mathbf{P}_{s,ur}^u$, we leverage the more confident prediction regions $\mathbf{P}_{t,mc}^u$ in pseudo labels, which tend to occur near high-density regions, to guide the optimization of less confident regions $\mathbf{P}_{s,lc}^u$ in unlabeled image predictions, which may appear at decision boundaries. $\mathbf{P}_{t,mc}^u = \mathbf{P}_{t,ur}^u \odot \mathcal{M}^{g}$, $\mathbf{P}_{s,lc}^u = \mathbf{P}_{s,ur}^u \odot \mathcal{M}^{g}$.

We employ L2 loss function to guide the optimization of low-confidence regions, while gradients from high-confidence regions are detached from back-propagation at this stage: $\mathcal{L}^g = \ell_{l2}(\mathbf{P}_{s,lc}^u, f_{detach}(\mathbf{P}_{t,mc}^u))$, where $\ell_{l2}$ represents the L2 loss function, and $f_{detach}(\cdot)$ denotes the detachment of gradients for back-propagation.

\subsection{Multi-scale mixing up strategy}
To encourage the model to perceive diverse scales of cardiac anatomical structures and to learn comprehensive common semantics from both labeled and unlabeled images, we implemented multi-scale mixing up strategy. First, we partition $\mathbf{X}^l$ and $\mathbf{X}^u$ into puzzle patches $\left \{ \mathbf{X}^l_j \right \}^{N^2}_{j=1}$ and $\left \{ \mathbf{X}^u_j \right \}^{N^2}_{j=1}$, respectively, where $\mathbf{X}^l_j,\mathbf{X}^u_j\in{\mathbb{R}^{W/N\times{H/N\times{C}}}}$, $N^2$ represents the total number of patches, and $j$ denotes the relative position of the patch in the image puzzle. We define the partition operation as $\mathcal{O}_{part}(\cdot)$. During the training phase, $N$ can be configured as a variable value, whereby different values of $N$ enable the model to perceive cardiac anatomical structures at different scales. Subsequently, puzzle patches $\mathbf{X}^l_j$ and $\mathbf{X}^u_j$ are randomly mixed up while maintaining their original positions, resulting in the creation of two new mixed images, denoted as $\mathbf{X}^{mix}_a,\mathbf{X}^{mix}_b\in{\mathbb{R}^{W\times{H\times{C}}}}$. The operation of mixing up puzzle patches is denoted as $\mathcal{O}_{mix}(\cdot)$. After being sequentially processed by the student network $\mathcal{F}_s(\cdot;\mathbf{\Theta}_s)$ and the softmax layer $\sigma(\cdot)$, mixed images yield predicted maps $\mathbf{P}^{mix}_a,\mathbf{P}^{mix}_b\in{\mathbb{R}^{K\times{W\times{H}}}}$. Next, employing the same mixing up method as before, we mix up ground-truth $\mathbf{Y}^l\in{\left \{ 0,1,...,K \right \}^{W\times{H}}}$ and pseudo-label $\mathbf{\widehat{Y}}^u$ to obtain mixed labels $\mathbf{Y}^{mix}_a,\mathbf{Y}^{mix}_b\in{\left \{ 0,1,...,K \right \}^{W\times{H}}}$ for supervising $\mathbf{P}^{mix}_a$ and $\mathbf{P}^{mix}_b$, respectively. The whole process is expressed succinctly as:
\begin{equation}
\mathbf{P}^{mix}_a,\mathbf{P}^{mix}_b = \sigma(\mathcal{F}_s(\mathcal{O}_{mix}(\mathcal{O}_{part}(\mathbf{X}^l,\mathbf{X}^u));\mathbf{\Theta}_s)),
\end{equation}
\begin{equation}
\mathbf{Y}^{mix}_a,\mathbf{Y}^{mix}_b = \mathcal{O}_{mix}(\mathcal{O}_{part}(\mathbf{X}^l,\mathbf{\widehat{Y}}^u)).
\end{equation}

The calculation of loss for $\mathbf{X}^{mix}_a$ and $\mathbf{X}^{mix}_b$ is denoted as:
\begin{equation}
\mathcal{L}^a = \ell_{ce}(\mathbf{P}^{mix}_a,\mathbf{Y}^{mix}_a) + \ell_{dice}(\mathbf{P}^{mix}_a,\mathbf{Y}^{mix}_a),
\end{equation}
\begin{equation}
\mathcal{L}^b = \ell_{ce}(\mathbf{P}^{mix}_b,\mathbf{Y}^{mix}_b) + \ell_{dice}(\mathbf{P}^{mix}_b,\mathbf{Y}^{mix}_b).
\end{equation}
where $\ell_{ce}(\cdot),\ell_{dice}(\cdot)$ denote multi-class Cross-Entropy and Dice loss functions respectively.

\subsection{Training strategy}
During the training phase, the parameters $\mathbf{\Theta}_s$ of the student network $\mathcal{F}_s(\cdot;\mathbf{\Theta}_s)$ are optimized by stochastic gradient descent (SGD), while the parameters $\mathbf{\Theta}_t$ of the teacher network $\mathcal{F}_t(\cdot;\mathbf{\Theta}_t)$ are updated from $\mathbf{\Theta}_s$ with exponential moving average (EMA)~\cite{tarvainen2017mean}. $\mathbf{\Theta}_s$ update by SGD with loss function at each iteration: $\mathcal{L}^{all} = (\mathcal{L}^a + \mathcal{L}^b) / 2 + \alpha\mathcal{L}^{rec} + \beta\mathcal{L}^{g}$, where $\alpha$ and $\beta$ are parameters for harmonizing the significance of each loss function. $\mathbf{\Theta}^{(k+1)}_t$ are updated at the ($k+1$)th iteration: $\mathbf{\Theta}^{(k+1)}_t = \lambda{\mathbf{\Theta}^{(k)}_t} + (1-\lambda)\mathbf{\Theta}^{(k)}_s$, where $\lambda$ represents the smoothing coefficient parameter.

\section{Experiments and results}
\subsection{Experimental setup}
\subsubsection{Datasets}
We evaluated our method on two echocardiography datasets: the public CAMUS (Cardiac Acquisitions for Multi-structure Ultrasound Segmentation) dataset~\cite{leclerc2019deep}  and a private clinical dataset. The CAMUS dataset comprises 3000 echocardiography images (four- and two-chamber views) from 500 patients at the University Hospital of St Etienne, France, aimed at segmenting cardiac substructures like the left ventricle endocardium, epicardium, and left atrium borders. The private dataset consists of 5671 echocardiography images (apical four-chamber view, low parasternal four-chamber view, and parasternal short-axis view of large artery) from 676 patients at Shanghai Children’s Medical Center (P.R.China). Its purpose is to segment the four cardiac chambers: right ventricle, left ventricle, right atrium, and left atrium. This study received ethics approval from the Shanghai Children's Medical Center Ethics Committee (Approval No. SCMCIRB-W2021058).

\subsubsection{Evaluation metrics}
We evaluate the performance of our method quantitatively by 4 common evaluation metrics follow in BCP~\cite{bai2023bidirectional}: Dice(\%), Jaccard(\%), the 95\% Hausdorff Distance (95HD)(pixel), and the average surface distance (ASD)(pixel).

\subsubsection{Implementation details}
We employ U-Net as the backbone of the architecture, with input image size set to $256\times256$ pixels. The parameter N for image puzzle partitions is randomly set to 2 or 3, dividing the image into 4 or 9 small puzzle patches. $\alpha$ and $\beta$ in the loss function are both set to 0.01. All experiments, using PyTorch and five-fold cross-validation, are conducted on Nvidia RTX 3080 GPU with SGD optimizer for model optimization.

\subsection{Quantitative evaluation}

\begin{table}
    \caption{Quantitative comparison results on the CAMUS dataset. (The unit of labeled data in the table is patients. U-Net is trained only on labeled data using supervised learning, whereas other semi-supervised methods are trained on both labeled and unlabeled data.)}
    \label{quantitative_camus}
    \begin{tabular}{ c | c | c  c  c  c}
        \hline
        Method & Labeled & Dice$\uparrow$ & Jaccard$\uparrow$ & 95HD$\downarrow$ & ASD$\downarrow$ \\
        \hline
        U-Net   & 5(1\%)   & 84.99 & 75.01 & 15.02 & 4.97 \\
        U-Net   & 25(5\%)  & 88.65 & 80.28 & 8.62 & 3.08 \\
        U-Net   & 500(All) & 90.67 & 83.40 & 6.84 & 2.47 \\
        \hline
        UA-MT\cite{yu2019uncertainty} & \multirow{7}*{5(1\%)} & 84.46 & 74.23 & 14.26 & 5.10 \\
        URPC\cite{luo2021efficient} & & 85.22 & 75.25 & 12.35 & 4.29 \\
        MC-Net\cite{wu2021semi} & & 85.54 & 75.76 & 12.69 & 4.32 \\
        CNN\_ViT\cite{wang2022cnn} & & 85.36 & 75.44 & 12.03 & 4.38 \\
        BCP\cite{bai2023bidirectional} & & 87.26 & 78.18 & 9.20 & 3.30 \\
        DCNet\cite{chen2023decoupled} & & 86.69 & 77.50 & 9.12 & 3.29 \\
        Ours & & \textbf{88.28} & \textbf{79.21} & \textbf{8.94} & \textbf{3.18} \\
        \hline
        UA-MT\cite{yu2019uncertainty} & \multirow{7}*{25(5\%)} & 88.47 & 80.03 & 8.40 & 3.04 \\
        URPC\cite{luo2021efficient} & & 88.78 & 80.51 & 7.84 & 2.84 \\
        MC-Net\cite{wu2021semi} & & 89.08 & 80.93 & 7.59 & 2.81 \\
        CNN\_ViT\cite{wang2022cnn} & & 88.24 & 79.57 & 9.45 & 3.39 \\
        BCP\cite{bai2023bidirectional} & & 89.26 & 81.24 & 7.45 & 2.74 \\
        DCNet\cite{chen2023decoupled} & & 89.21 & 81.30 & 7.44 & 2.75 \\
        Ours & & \textbf{89.63} & \textbf{81.75} & \textbf{7.22} & \textbf{2.64} \\
        \hline
\end{tabular}
\end{table}

\begin{table}
    \caption{Quantitative comparison results on the private clinical dataset. (Refer to Table~\ref{quantitative_camus} for the content in parentheses.)}
    \label{quantitative_private}
    \begin{tabular}{ c | c | c  c  c  c}
        \hline
        Method & Labeled & Dice$\uparrow$ & Jaccard$\uparrow$ & 95HD$\downarrow$ & ASD$\downarrow$ \\
        \hline
        U-Net   & 7(1\%)    & 69.50 & 59.73 & 62.88 & 9.80 \\
        U-Net   & 34(5\%)   & 76.52 & 68.32 & 61.79 & 4.64 \\
        U-Net   & 676(All)  & 78.51 & 72.04 & 75.97 & 2.15 \\
        \hline
        UA-MT\cite{yu2019uncertainty} & \multirow{7}*{7(1\%)} & 69.03 & 59.15 & 67.30 & 9.71 \\
        URPC\cite{luo2021efficient} & & 69.96 & 60.44 & 64.40 & 6.69 \\
        MC-Net\cite{wu2021semi} & & 70.54 & 61.02 & 65.24 & 7.77 \\
        CNN\_ViT\cite{wang2022cnn} & & 70.49 & 60.65 & 64.89 & 7.13 \\
        BCP\cite{bai2023bidirectional} & & 70.43 & 60.89 & 65.49 & 7.46 \\
        DCNet\cite{chen2023decoupled} & & 70.25 & 60.96 & 63.81 & 5.48 \\
        Ours & & \textbf{71.75} & \textbf{61.68} & \textbf{63.62} & \textbf{5.44} \\
        \hline
        UA-MT\cite{yu2019uncertainty} & \multirow{7}*{34(5\%)} & 76.49 & 68.29 & 61.92 & 4.69 \\
        URPC\cite{luo2021efficient} & & 75.95 & 67.78 & 61.60 & 4.10 \\
        MC-Net\cite{wu2021semi} & & 76.75 & 68.68 & 62.14 & 4.40 \\
        CNN\_ViT\cite{wang2022cnn} & & 76.90 & 68.95 & 62.06 & 4.36 \\
        BCP\cite{bai2023bidirectional} & & 77.08 & 68.85 & 62.32 & 4.69 \\
        DCNet\cite{chen2023decoupled} & & 77.33 & 69.14 & 61.76 & 4.50 \\
        Ours & & \textbf{77.90} & \textbf{69.95} & \textbf{61.44} & \textbf{4.03} \\
        \hline
\end{tabular}
\end{table}

\begin{figure}[ht]
    \centering
    \includegraphics[width=0.5\textwidth]{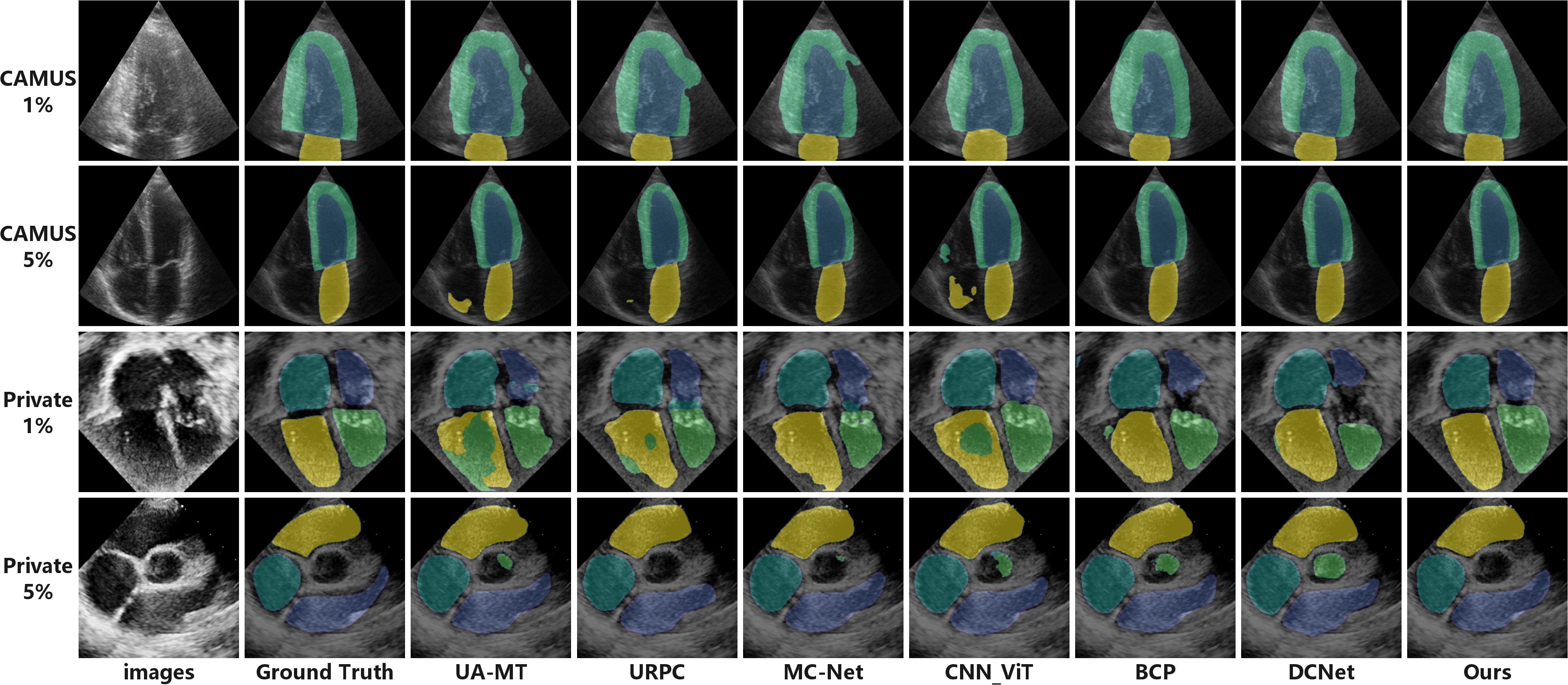}
    \caption{Visualization results of comparative experiments between our method and others on the CAMUS and private clinical dataset with 1\% and 5\% labeled ratios.} \label{compare}
\end{figure}

We compare our method with existing ones: UA-MT~\cite{yu2019uncertainty}, URPC~\cite{luo2021efficient}, MC-Net~\cite{wu2021semi}, CNN\_ViT~\cite{wang2022cnn}, BCP~\cite{bai2023bidirectional}, and DCNet~\cite{chen2023decoupled}. Results from other compared methods were obtained using the same experimental setting in BCP~\cite{bai2023bidirectional} for fair comparisons. Table~\ref{quantitative_camus} shows the quantitative performance of segmentation results on CAMUS and with 1\% and 5\% labeled ratios. Our approach outperforms other competitors under the condition of extremely limited labeled images (1\%), achieving a 1.02\% improvement in Dice score on the private dataset. Furthermore, our method achieves the best performance in the other three metrics. Table~\ref{quantitative_private} shows the quantitative performance of segmentation results on private clinical dataset with 1\% and 5\% labeled ratios. Our approach continues to outperform other approaches under the condition of (1\%) labeled images, achieving a 1.21\% improvement in Dice score on the private clinical dataset. This implies that reflecting upon and enhancing the quality of pseudo-label generation is advantageous for improving the performance and robustness of the network. Surprisingly, the performance of UA-MT appears to be even worse than the lower bound. It is possible that UA-MT failed to address challenges posed by characteristics of echocardiography and failed to generate high-quality pseudo-labels, and thus misled the student network. Visualization and additional analysis are shown in the supplementary material.

Moreover, we present the visualization of segmentation results in Fig.~\ref{compare}. As shown in the figure, our method achieves a more delicate edge segmentation. Some methods erroneously identify the interior of larger objects (e.g., right ventricle) in certain images as belonging to other categories. Some other methods mistakenly classify unnecessary segmentation areas in some images as a specific object (e.g., aorta).

\subsection{Ablation studies}

\begin{table}[ht]
    \caption{The ablation study results for the two major strategies within the proposed framework. (\textbf{ERS} denotes Error Reflection Strategy, \textbf{MMS} represents Multi-scale Mixing up Strategy, and \textbf{L} denotes Labeled data.)}
    \label{ablation_all}
    \begin{tabular}{ c c | c | c  c | c | c  c}
        \hline
        ERS & MMS & L & Dice$\uparrow$ & 95HD$\downarrow$ & L & Dice$\uparrow$ & 95HD$\downarrow$ \\
        \hline
        & & \multirow{4}*{1\%} & 85.34 & 12.56 & \multirow{4}*{5\%} & 88.92 & 7.81 \\
        \checkmark & & & 86.58 & 9.28 & & 89.11 & 7.57 \\
        & \checkmark & & 87.29 & 9.15 & & 89.24 & 7.46 \\
        \checkmark & \checkmark & & \textbf{88.28} & \textbf{8.94} & & \textbf{89.63} & \textbf{7.22} \\
        \hline
\end{tabular}
\end{table}

To validate the effectiveness of our designs, we conduct a series of ablation studies on public dataset CAMUS. First, we evaluate the effectiveness of the error reflection strategy and the multi-scale mixing up strategy in our proposed framework. As shown in Table~\ref{ablation_all}, the experimental results demonstrate that the proposed strategies have all contributed to performance gains.

\begin{table}[ht]
    \caption{The ablation study results for the error reflection strategy. (\textbf{All} denotes our complete method, \textbf{-} represents removal, \textbf{S1} indicates the reflective reconstruction step of the error reflection strategy, \textbf{S2} denotes the guidance correction step of the error reflection strategy, and \textbf{AS} represents auxiliary sketch.)}
    \label{ablation_reflection}
    \begin{tabular}{ c | c | c  c | c | c  c}
        \hline
        Method & Labled & Dice$\uparrow$ & 95HD$\downarrow$ & Labled & Dice$\uparrow$ & 95HD$\downarrow$ \\
        \hline
        All-AS & \multirow{4}*{5(1\%)} & 85.67 & 10.12 & \multirow{4}*{25(5\%)} & 88.86 & 7.91 \\
        All-S1  & & 87.64 & 9.11 & & 89.39 & 7.42 \\
        All-S2 & & 87.96 & 9.02 & & 89.52 & 7.34 \\
        All & & \textbf{88.28} & \textbf{8.94} & & \textbf{89.63} & \textbf{7.22} \\
        \hline
\end{tabular}
\end{table}

\begin{figure}[ht]
    \includegraphics[width=0.5\textwidth]{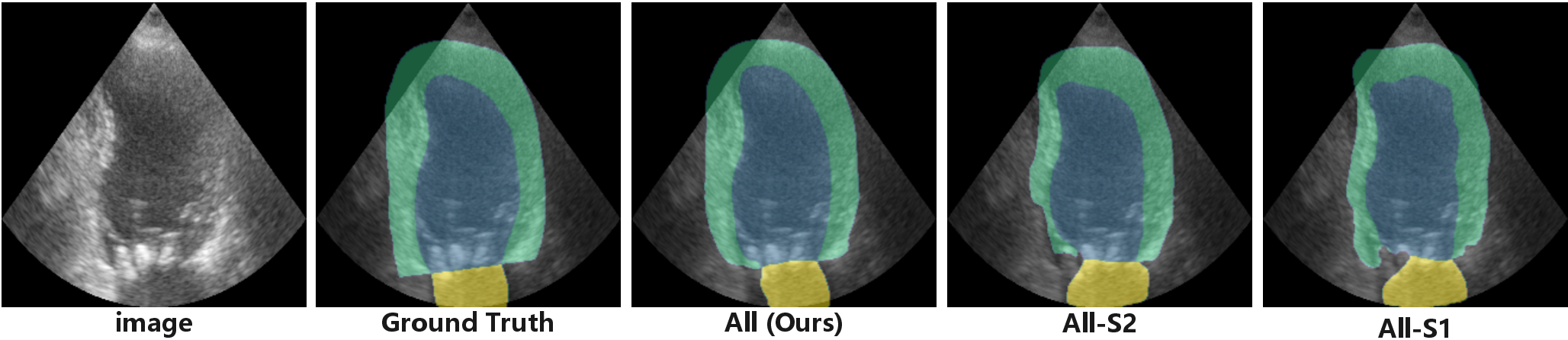}
    \caption{Visualizing the quality of pseudo-label generation under varying conditions in error reflection strategy ablation studies. (Refer to Table~\ref{ablation_reflection} for the explanation of \textbf{All}, \textbf{-}, \textbf{S1} and \textbf{S2})} \label{vis}
\end{figure}

We focus our study on examining whether error reflection strategy can overcome the challenges in pseudo-label generation for semi-supervised echocardiography segmentation posed by the poor contrast, blurred edge details and noise characteristics of echocardiography. In one experiment, we remove the reconstruction reflection step, replace the process of generating the unreliable region map with a simple Softmax function evaluation, and retain the guidance correction step. In another experiment, we retain the reconstruction reflection step but remove the guidance correction step. As shown in Table~\ref{ablation_reflection}, the experimental results indicate that both steps of the error reflection strategy are indispensable for achieving optimal performance. Additionally, we examine the significance of auxiliary sketches: the generation of distorted proxies without auxiliary sketches adversely impacts segmentation quality. As illustrated in Fig.~\ref{vis}, each of our designs contributes to performance enhancement, effectively improving the generation quality of pseudo-labels.

\begin{table}[ht]
    \caption{The ablation study results for the error reflection strategy. ($N$ denotes number of puzzle patches, \textbf{2/3} indicates randomly setting $N$ to either 2 or 3.)}
    \label{ablation_puzzle}
    \begin{tabular}{ c | c | c  c | c | c  c}
        \hline
        $N$ & Labled & Dice$\uparrow$ & 95HD$\downarrow$ & Labled & Dice$\uparrow$ & 95HD$\downarrow$ \\
        \hline
        2 & \multirow{4}*{5(1\%)} & 87.39 & 9.07 & \multirow{4}*{25(5\%)} & 89.21 & 7.45 \\
        3  & & 87.67 & 9.10 & & 89.32 & 7.38 \\
        4 & & 84.75 & 14.52 & & 88.48 & 8.41 \\
        2/3 & & \textbf{88.28} & \textbf{8.94} & & \textbf{89.63} & \textbf{7.22} \\
        \hline
\end{tabular}
\end{table}

Furthermore, we study the impact of number (N) of puzzle patches. As shown in Table~\ref{ablation_puzzle}, the experimental results reveal that the model performs optimally when N is randomly assigned as either 2 or 3. This suggests that the strategy can effectively improve empirical mismatch between labeled and unlabeled data distribution, and facilitate the perception of cardiac anatomical structures at different scales, thereby enhancing performance.

\section{Conclusion}
In this paper, we reflect upon the characteristics of poor contrast, blurred edge details and noise in echocardiography, and propose an error reflection strategy for semi-supervised segmentation of echocardiography. The strategy, consisting of a reconstruction reflection step and a guidance correction step, aims to improve the quality of pseudo-label generation, thereby improving the model performance. Additionally, we introduce a data augmentation strategy, namely multi-scale mixing up strategy, to reduce the distribution mismatch between labeled and unlabeled images and facilitate the perception of various scales of cardiac anatomical structures. Experimental results on a public echocardiography dataset and a private clinical echocardiography dataset verify the superiority of the proposed method.

\section*{Acknowledgment}
We thank Qiming Huang, a Ph.D. student from the University of Birmingham, for insights. This work was partially supported by the National Natural Science Foundation of China (Grant No. 62071285), Sanya Science and Technology Innovation Program (Grant No. 2022KJCX41), and Pediatric Medical Consortium Scientific Research Project of Shanghai Children's Medical Center affiliated to Shanghai Jiao Tong University School of Medicine.

\bibliographystyle{IEEEtranS}
\bibliography{ref.bib}

\end{document}